\documentstyle[epsfig,rotate,twoside,fleqn,espcrc2]{article}

\newcommand{\AmS}{{\protect\the\textfont2
  A\kern-.1667em\lower.5ex\hbox{M}\kern-.125emS}}

\title{Baryon mass extrapolation}

\author{Derek B. Leinweber, Anthony W. Thomas, Kazuo Tsushima and
        Stewart V. Wright\address{Department of Physics and
        Mathematical Physics and
        Special Research Centre for the Subatomic Structure of
        Matter,
        University of Adelaide, Australia 5005}
        \thanks{
        Supported by the Australian Research Council.
        }}
       
\begin{document}

\begin{abstract}
Consideration of the analytical properties of pion-induced baryon
self-energies leads to new functional forms for the extrapolation of
light baryon masses.  These functional forms reproduce the leading
non-analytic behavior of chiral perturbation theory, the correct
heavy-quark limit and have the advantage of containing information on
the extended structure of hadrons.  The forms involve only three
unknown parameters which may be optimized by fitting to present
lattice data.  Recent dynamical fermion results from CP-PACS and
UK-QCD are extrapolated using these new functional forms.  We also use
these functions to probe the limit of the chiral perturbative regime
and shed light on the applicability of chiral perturbation theory to
the extrapolation of present lattice QCD results.
\vspace{1pc}
\end{abstract}

\maketitle

\section{FORMALISM}

In recent years there has been tremendous progress in the computation
of baryon masses within lattice QCD.  Still, it remains necessary to
extrapolate the calculated results to the physical pion mass ($\mu
=140$ MeV) in order to make a comparison with experimental data.  In
doing so one necessarily encounters some non-linearity in the quark
mass (or $m_{\pi }^{2}$), including the non-analytic behavior
associated with dynamical chiral symmetry breaking.  We recently
investigated this problem for the case of the nucleon magnetic moments
\cite{Leinweber:1998ej}.  It is vital to develop a sound understanding
of how to extrapolate to the physical pion mass.

\subsection{Self-Energy Contributions}

\begin{figure}[t]
\centering{\
\epsfig{file=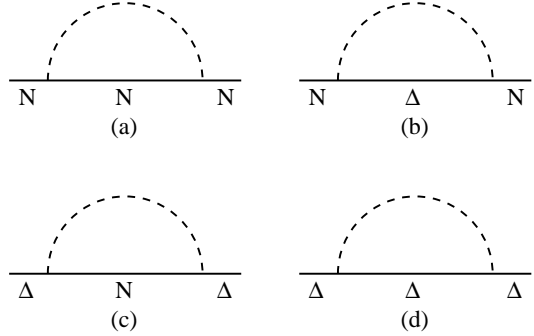,width=7cm}}
\vspace{-24pt}
\caption{One-loop pion induced self-energy of the nucleon and the
delta.
\label{SE-fig}}
\end{figure}

Chiral symmetry is dynamically broken in QCD and the pion alone is a
near Goldstone boson.  It is strongly coupled to baryons and plays a
significant role in $N$ and $\Delta$ self-energies.  The one-loop pion
induced self-energies of the $N$ and $\Delta$ are given by the
processes shown in Fig.~\ref{SE-fig}.

In the standard heavy baryon limit, the analytic expression for the
pion cloud correction to the masses of the $N$ and $\Delta$ are of the
form \cite{Leinweber:1999ig}
\begin{equation}
\delta M_{N}=\sigma _{NN}+\sigma _{N\Delta }\, ; \quad
\delta M_{\Delta }=\sigma _{\Delta \Delta }+\sigma _{\Delta N}\, ,
\label{Init-ND-eqn}
\end{equation}
where
\begin{equation}
\sigma _{NN} = \sigma_{\Delta \Delta} = 
-\frac{3\, g_{A}^{2}}{16\, \pi ^{2}f^{2}_{\pi }}
\int _{0}^{\infty }dk\frac{k^{4}u^{2}_{NN}(k)}{w^{2}(k)}\, ,
\label{SE-NN-integral-eqn}
\end{equation}
\begin{equation}
\sigma _{N\Delta } =  -\frac{6\, g_{A}^{2}}{25\, \pi ^{2}f^{2}_{\pi }}
\int ^{\infty }_{0}dk\frac{k^{4}u^{2}_{N\Delta }(k)}{w(k)(\Delta
M+w(k))}\, ,
\label{SE-ND-integral-eqn} 
\end{equation}
\begin{equation}
\sigma _{\Delta N} = \frac{3\, g_{A}^{2}}{50\, \pi^{2}f^{2}_{\pi }}
\int _{0}^{\infty }dk\frac{k^{4}u^{2}_{N\Delta }(k)}
{w(k)(\Delta M-w(k))}\, .
\label{SE-DN-integral-eqn} 
\end{equation}
Here $\Delta M=M_{\Delta }-M_{N}$ , $g_{A}=1.26$ is the axial
charge of the nucleon, $w(k)=\sqrt{k^{2}+m^{2}_{\pi }}$ is the pion
energy and $u_{NN}(k)$, $u_{N\Delta }(k)$, $\ldots$ are the $NN\pi$,
$N\Delta \pi$, $\ldots$ form factors associated with the emission of a
pion of three-momentum $k$.  The form factors reflect the finite size
of the baryonic source of the pion field and suppress the emission
probability at high virtual pion momentum.  As a result, the
self-energy integrals are not divergent.

The leading non-analytic (LNA) contribution of these self-energy
diagrams is associated with the infrared behavior of the corresponding
integrals; i.e.\ the behavior as $k\rightarrow 0$.  As a consequence,
the leading non-analytic behavior does not depend on the details of
the form factors.  Indeed, the well known results of chiral
perturbation theory \cite{HBchiPT,Lebed94} are reproduced even when
the form factors are approximated by $u(k) = \theta(\Lambda - k)$.  

Of course, our concern with respect to lattice QCD is not so much the
behavior as $m_{\pi }\rightarrow 0$, but the extrapolation from high
pion masses to the physical pion mass.  In this context the branch
point at $m_{\pi }^{2}=\Delta M^{2}$, associated with transitions of
$N \to \Delta$ or $\Delta \to N$, is at least as important as the LNA
behavior near $m_{\pi }=0$.

Heavy quark effective theory suggests that as $m_{\pi }\rightarrow
\infty$ the quarks become static and hadron masses become proportional
to the quark mass.  In this spirit, corrections are expected to be of
order $1/m_{q}$ where $m_{q}$ is the heavy quark mass.  The presence
of a cut-off associated with the form factor acts to suppress the pion
induced self energy for increasing pion masses, as evidenced by the
$m_{\pi }^{2}$ in the denominators of Eqs. (\ref{SE-NN-integral-eqn}),
(\ref{SE-ND-integral-eqn}) and (\ref{SE-DN-integral-eqn}).  While some
$m_{\pi }^{2}$ dependence in the form factor is expected, this is a
second-order effect and does not alter the qualitative feature of the
self-energy corrections tending to zero as $1/m_{\pi }^{2}$ in the
heavy quark limit.

Rather than simplifying our expressions to just the LNA terms, we
retain the complete expressions \cite{Leinweber:1999ig}, as they
contain important physics that would be lost by making a
simplification.  We note that keeping the entire form is not in
contradiction with $\chi$PT.  However, as one proceeds to larger quark
masses, differences between the full forms and the expressions in the
chiral limit will become apparent, highlighting the importance of the
branch point and the form factor reflecting the finite size of
baryons.

As a result of these considerations, we propose to use the analytic
expressions for the self-energy integrals corresponding to a sharp
cut-off in order to incorporate the correct LNA structure in a simple
three-parameter description of the $m_{\pi }$ dependence of the $N$
and $\Delta$ masses.  In the heavy quark limit hadron masses become
proportional to the quark mass.  Hence we can simulate a linear
dependence of the baryon masses on the quark mass, $m_q$, in this
region, by adding a term involving $m^{2}_{\pi }$.  The functional
form for the mass of the nucleon suggested by this analysis is then:
\begin{equation}
\label{N-form-eqn}
M_{N}=\alpha_{N} + \beta_{N}m_{\pi }^{2} + \sigma_{NN}(\Lambda_N
) + \sigma_{N\Delta }(\Lambda_N )\, ,
\end{equation}
while that for the $\Delta$ is:
\begin{equation}
\label{D-form-eqn}
M_{\Delta }=\alpha_{\Delta } + \beta_{\Delta } m_{\pi }^{2} +
\sigma_{\Delta \Delta }(\Lambda_\Delta ) + \sigma_{\Delta
N}(\Lambda_\Delta )\, .  
\end{equation}

\subsection{Model Dependence}

\begin{figure}[t]
\centering{\
\rotate{\epsfig{file=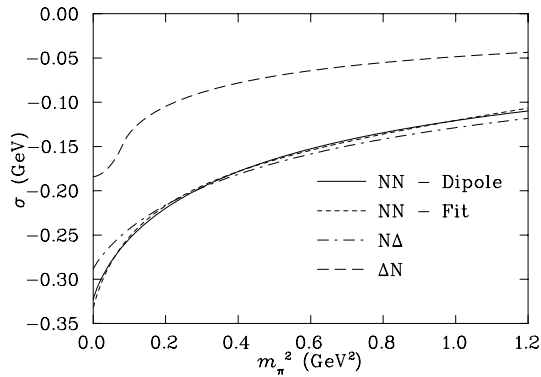,height=7cm}}}
\vspace{-24pt}
\caption{ The self-energy contribution $\sigma_{NN}$ for a 1 GeV
dipole form factor (solid curve) is compared with a sharp cut-off form
factor $\theta(\Lambda_N - k)$ (fine-dash curve).  Self-energy
contributions $\sigma_{N \Delta}$ (dot-dash) and $\sigma_{\Delta N}$
(long-dash) for a 1 GeV dipole are also illustrated.
\label{SE-Fit}}
\end{figure}

The use of a sharp cut-off, $u(k) = \theta(\Lambda - k)$, as a form
factor may seem somewhat unfortunate given that phenomenology suggests
a dipole form factor better approximates the axial-vector form factor.
However, the sensitivity to such model-dependent issues is shown to be
negligible in Fig.~\ref{SE-Fit}.  There, the self-energy contribution
$\sigma_{NN} (= \sigma_{\Delta \Delta})$ for a 1 GeV dipole form
factor (solid curve) is compared with a sharp cut-off form factor
combined with the standard $\alpha+\beta\, m_{\pi }^{2}$ terms of
(\ref{N-form-eqn}) or (\ref{D-form-eqn}).  Optimizing $\Lambda$,
$\alpha$ and $\beta$ provides the fine-dash curve of
Fig.~\ref{SE-Fit}.  Differences are at the few MeV level indicating
negligible sensitivity to the actual analytic structure of the form
factor.

Here we have focused on the pion self-energy contribution to the $N$
and $\Delta$ form factors.  Only the pion displays a rapid mass
dependence as the chiral limit is approached.  Other mesons
participating in similar diagrams do not give rise to such rapidly
changing behavior and can be accommodated in the
$\alpha+\beta\, m_{\pi }^{2}$ terms of (\ref{N-form-eqn}) or
(\ref{D-form-eqn}).  Moreover, the form factor suppresses the
contributions from more massive intermediate states including multiple
pion dressings.  Other multi-loop pion contributions renormalize the
vertex and hence we use the renormalized coupling $g_A$ as a measure of
the pion-nucleon coupling.

\section{ANALYSIS}

We consider two independent dynamical-fermion lattice simulations of
the $N$ and $\Delta$ masses.  We select results from CP-PACS's
\cite{CP-PACSlight} $12^3 \times 32$ and $16^3 \times 32$ simulations
at $\beta=1.9$, and UKQCD's \cite{Allton:1998gi} $12^3 \times 24$
simulations at $\beta=5.2$.  

\begin{figure}[t]
\centering{\
\rotate{\epsfig{file=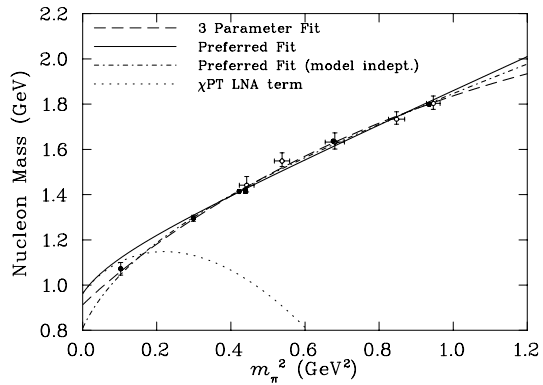,height=7cm}}
\vspace{-24pt}
\caption{A comparison of phenomenological fitting functions for the
mass of the nucleon.  The solid curve corresponds to our preferred fit
of the functional form of (\ref{N-form-eqn}) with $\Lambda$
constrained to reproduce a 1 GeV dipole form factor.  The dash-dot
curve illustrates the unconstrained fit.  The three parameter fit
(dashed) corresponds to letting $\gamma$ of (\protect\ref{Naive-eqn})
vary as an unconstrained fit parameter.  The dotted curve corresponds
to (\protect\ref{Naive-eqn}) with $\gamma$ set equal to the value
known from $\chi$PT.  The lattice data from are CP-PACS (solid) and
UKQCD (open), each with a 5\% scale change to provide consistency.
\label{3fits-N-fig}}}
\end{figure}

Figure \ref{3fits-N-fig} displays fits of (\ref{N-form-eqn}) to the
lattice data.  In order to perform fits in which $\Lambda$ is
unconstrained, it is essential to have lattice simulations at light
quark masses approaching $m^{2}_{\pi }\sim 0.1$ GeV$^{2}$.  

It is common to see the use of the following $\chi$PT-motivated
expression for the mass dependence of hadron masses,
\begin{equation}
\label{Naive-eqn}
M_{N}=\alpha +\beta m_{\pi }^{2}+\gamma m_{\pi }^{3} \, .
\end{equation}
The result of such a fit for the $N$ is shown as the dashed curve in
Fig. \ref{3fits-N-fig}. The coefficient of the $m_{\pi }^{3}$ term in
a three parameter fit is $-0.761$.  This disagrees with the
coefficient of $-5.60$ known from $\chi$PT (which is correctly
incorporated in (\ref{N-form-eqn}) and illustrated as the solid and
dash-dot curves of Fig. \ref{3fits-N-fig}) by almost an order of
magnitude.  This clearly indicates the failings of (\ref{Naive-eqn}).

The dotted curve of Fig.~\ref{3fits-N-fig} indicates the leading
non-analytic term of the chiral expansion dominates from the chiral
limit up to the branch point at $m_{\pi}=\Delta M \simeq 300$ MeV,
beyond which $\chi$PT breaks down.  The curvature around $m_{\pi
}=\Delta M$, neglected in previous extrapolations of the lattice data,
leads to shifts in the extrapolated masses of the same order as the
departure of lattice estimates from experimental measurements.

\end{document}